# Explosive Formation and Dynamics of Vapor Nanobubbles around a Continuously Heated Gold Nanosphere

Lei Hou, Mustafa Yorulmaz, Nico R Verhart, Michel Orrit*

LION, Huygens-Kamerlingh Onnes Laboratory, Leiden University

Niels Bohrweg 2, 2300RA Leiden (Netherlands)

*Abstract*

*We form sub-micrometer-sized vapor bubbles around a single laser-heated gold nanoparticle in a liquid and monitor them through optical scattering of a probe laser. Bubble formation is explosive even under continuous-wave heating. The fast, inertia-governed expansion is followed by a slower contraction and disappearance after some tens of nanoseconds. In a narrow range of illumination powers, bubble time traces show a clear echo signature. We attribute it to sound waves released upon the initial explosion and reflected by flat interfaces, hundreds of microns away from the particle. Echoes can trigger new explosions. A nanobubble's steady state (with a vapor shell surrounding the heated nanoparticle) can be reached by a proper time profile of the heating intensity. Stable nanobubbles could have original applications for light modulation and for enhanced optical-acoustic coupling in photoacoustic microscopy.*

* corresponding author

1. Introduction

Gas bubbles in liquids are involved in many processes and applications [1]. Cavitation bubbles cause mechanical damage [2], and even can produce high-energy electromagnetic radiation, an effect called sonoluminescence [3]. The dynamics of bubbles driven by acoustic waves is highly nonlinear and displays many complex phenomena including chaos [1]. In the present work, we consider nanobubbles, i.e., bubbles with diameters of a few tens to hundreds of nanometers. Nanobubbles can be generated by heating a metal (gold) nanoparticle in a liquid with a focused laser pulse. Compared to earlier experiments on microbubbles (a few microns to tens of microns in diameter), nanobubbles are more difficult to observe, study and manipulate. However, they may give rise to simpler or different properties, for example because liquid flows around nanobubbles present lower Reynolds numbers [4].

The standard method to form a gas bubble optically is to illuminate an absorbing liquid with a (sub-)picosecond laser pulse [5,6]. Alternatively, the energy absorbed by a metal nanoparticle is transferred to the liquid by conduction. Because of the fast excitation and of the high heat conductivity of the metal, the particle's temperature is raised within a few picoseconds to well above the boiling temperature of the liquid, whose sudden vaporization leads to a necessarily explosive expansion of hot steam, pushing the liquid away and launching bubble dynamics in the nanosecond time regime. Transient nanobubbles produced by short laser pulses have been monitored [7,8,9] i) by optical imaging with short pulses, giving direct access to bubble size, ii) through the time dependence of probe light scattered off the bubble, giving time-resolved information, iii) by acoustic detection of emitted sound waves [7], or by the combination of electric conductivity through a nanopore and optical detection [10]. The environment change upon boiling can be detected via the particle's plasmon resonance [11], or by photothermal detection [12] but with limited time resolution. Ensembles of nanobubbles were studied by light scattering and by small-angle X-ray scattering of short X-ray pulses [13], or by femtosecond pump-probe spectroscopy [14,15], but these methods do not apply to single nanobubbles. Baffou et al. [16] used an imaging microscope to create microbubbles with continuous illumination of a single gold nanoparticle, also with low time resolution. Recently, Cichos' group modulated a probe beam by an isotropic bubble in a nematic liquid crystal close to its phase transition [17], but this process is much slower than the liquid-vapor equilibria considered here.

The aim of the present work was to produce a single stable nanobubble that could be studied optically. Stable nanobubbles are interesting for their fundamental properties [18,19] and may have useful applications [20]. A nanobubble is an efficient two-way transducer between acoustic and optical waves because of its large compressibility and optical scattering cross-section. A stable nanobubble could be an attractive beacon, source, or detector in photoacoustic microscopy [21]. Thus, we set out to heat a gold nanoparticle with a continuous laser beam, striving towards a precise balance between heat production in the nanoparticle and heat loss to the surrounding liquid. We monitor bubble formation and dynamics optically, which provides high time resolution down to the nanosecond, single-shot regime. Future experiments with short probe pulses and stroboscopic detection may give access to the sub-picosecond regime.

Let us briefly discuss the possible steady states for a heated metal nanosphere in a liquid such as water. Upon moderate heating, the particle is surrounded by a temperature gradient of hot water, as found for a gold nanorod in an optical trap [22]. We call this steady state regime I. Above a critical temperature, we expect to reach a second steady state, regime II, with a steam bubble around the nanoparticle. This steam bubble remains in equilibrium with the liquid if the vapor pressure balances the ambient pressure, increased by the Laplace pressure $p_L$ created by the liquid's surface tension $A(T)$, $T$ being the interface temperature. The Laplace pressure scales inversely with nanobubble radius $R$ according to $p_L(T) = \frac{2A(T)}{R}$, and reaches 30 atm for a water bubble with 100 nm diameter at room temperature. For water, $A(T = 300 \text{ K}) =$ 73 mN/m. Therefore, the water temperature must be much higher than 373 K (the macroscopic boiling temperature at ambient pressure) to form a bubble around a nanoparticle. Figure 1a shows a phase diagram of regimes I and II for a heated nanosphere with radius $r_{particle}$ in water, calculated with simple assumptions (see Supplementary Information). Figure 1b shows the temperature profile around a nanosphere ($r_{particle}$ = 40 nm) immersed in another liquid, n-pentane. The large temperature gradient in the vapor is due to its low heat conductivity. The dashed line in Fig.1b shows the boiling temperature when the gas bubble just touches the sphere and shows the boundary between regimes I and II for n-pentane. For very small particles, diameter<10 nm, the interface approaches the critical temperature (647 K for water, 469.8 K for n-pentane), and no clear interface appears anymore [23].

Here, we search for regime II of stable nanobubbles. To our surprise, we could not pass continuously from regime I to II upon increasing the heating. Instead, the system undergoes an explosive transition. After characterizing this instability, we show that a bubble can persist for a short time (about one microsecond in our current experiment) under a proper time profile of the heating intensity.

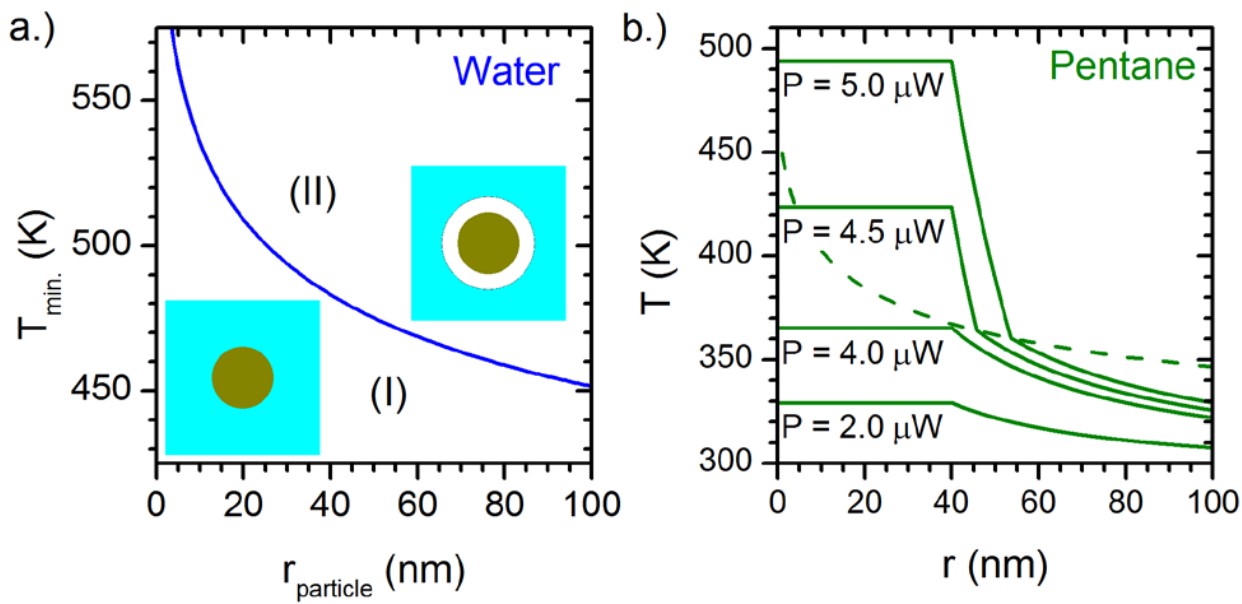

*Figure 1: Phase diagram and temperature profile around a heated particle in liquid. (a): Phase diagram of a heated particle in water. The line is the minimum temperature to reach the boiling as a function of particle radius. In steady state, a nanobubble develops around the particle above the boiling line. The boiling temperature is significantly shifted with respect to ambient pressure (373 K). (b): Radial temperature profile around a nanosphere (diameter 80 nm) calculated in pentane. The dashed line indicates the boiling phase transition as a function of particle size.*

2. Methods

One of our main concerns in this work has been the reproducibility of our experiments. We studied the same nanoparticle for a long time in a number of different experimental conditions for direct comparison. We therefore worked with gold nanospheres immobilized on a glass substrate, which can be imaged with high precision in a standard confocal microscope. To obtain large enough optical signals, the size of nanoparticles was chosen not too small, 80 nm in diameter. We also expect surface effects and irreversible shape changes of the particle or of the particle-substrate area to be (relatively) less important for large particles.

For the sample preparation, we used gold nanospheres and immobilized them on a glass coverslip by spin coating. The particles and the glass substrates were cleaned from organic ligands by repeated flushing and ozone cleaning. We checked that all particles used in the present work were isolated single ones, by photothermal contrast [24]. The particles were covered with the liquid (water or n-pentane), so that bubbles could form in the half-space limited by the flat glass-liquid interface. This breaks the spherical symmetry of the nanobubble with respect to the gold nanoparticle and the interface (see the scheme in Fig.2.a), but the nanobubble itself could have a spherical shape. Our first experiments were done in water for convenience and because it is by far the most interesting fluid for applications. However, boiling water around a nanoparticle requires temperatures up to 550 K, and the temperature of the gold particle can easily reach several hundreds of Kelvin above the water's temperature (see Fig. 1b). Such high temperatures can lead to irreversible changes in particle shape (movement of facets and of surface atoms) and, more critically, to changes in the contact area between gold and glass. Therefore, to limit the possibility of such random or irreversible changes we used n-pentane in most experiments. Pentane has a low boiling point (309 K at atmospheric pressure) so that we estimated that the particle's temperature did not exceed 370 K upon boiling under our experimental conditions. This temperature remains low enough to neglect surface rearrangements even after long illumination times.

We investigate the nanobubble optically only. The advantages of an all-optical investigation of the nanobubble are its speed, non-invasiveness and sensitivity. The optical setup, shown in the Fig. 2.b and Supplementary Information, is a classical photothermal microscope [25]. Photothermal microscopy is a technique based on the absorption of small objects such as gold nanoparticles. A modulated heating beam heats the nanoparticle and creates a temperature gradient, or thermal lens, around the absorbing object. A non-resonant probe beam, which is spatially overlapped with the heating beam, is scattered by the thermal lens and interferes with a reference beam, usually the transmitted or reflected probe beam. The interfering probe beam is then collected by a photodetector such as a photodiode, and the signal is demodulated by lock-in amplifier. We used the photothermal signal to overlap the heating (532 nm) and probing (815 nm) beams, to identify single gold nanoparticles in the sample, and also to find the critical intensity required for boiling (see 3.1). However, the photothermal signal, being produced by a lock-in amplifier with an integration time of at least 0.1 ms, was too slow to follow fast bubble kinetics in the nanosecond and microsecond domains. For these time-resolved measurements, the probe signal collected in reflection mode in bright-field scattering [26] was directly fed into the fast photodiode and the electronic signal was recorded in a fast oscilloscope with large memory.

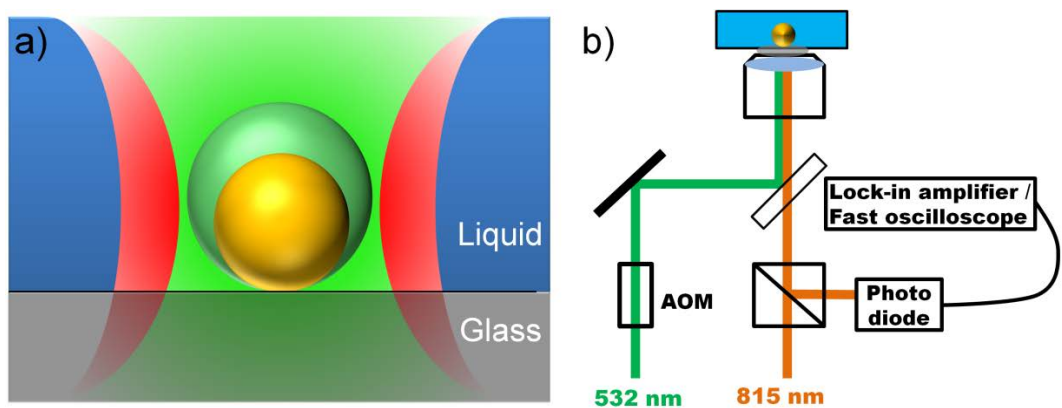

*Figure 2: Scheme of the vapor bubble formation and the experimental setup. a) a cartoon showing the asymmetric vapor bubble formation with respect to the nanosphere due to the glass-liquid interface, the gold nanosphere is heated continuously by the heating beam; b) scheme of the optical setup used here for nanobubble studies.*

3. Results and discussion

*3.1 Photothermal detection*

We started our study of single immobilized nanospheres with photothermal contrast (see Section 2). At low heating power, we only find continuous heating of the liquid around the particle. Above a critical pump intensity, however, the photothermal signal increases suddenly as shown in Fig.3, due to water boiling and nanobubble formation. Repeated heating cycles around the threshold power show fluctuations of the transition power by a few %. The particle temperature, estimated from the absorption cross section and the heat conductivity of water and glass (see Supplementary

Information), corresponds well with the simple model of Fig. 1. Yet, this steady-state model fails to explain the strong variability of the signal and the irreproducibility of successive transitions. Even at its highest time resolution (0.1 ms), the lock-in detection is too slow to follow the dynamics of the nanobubble.

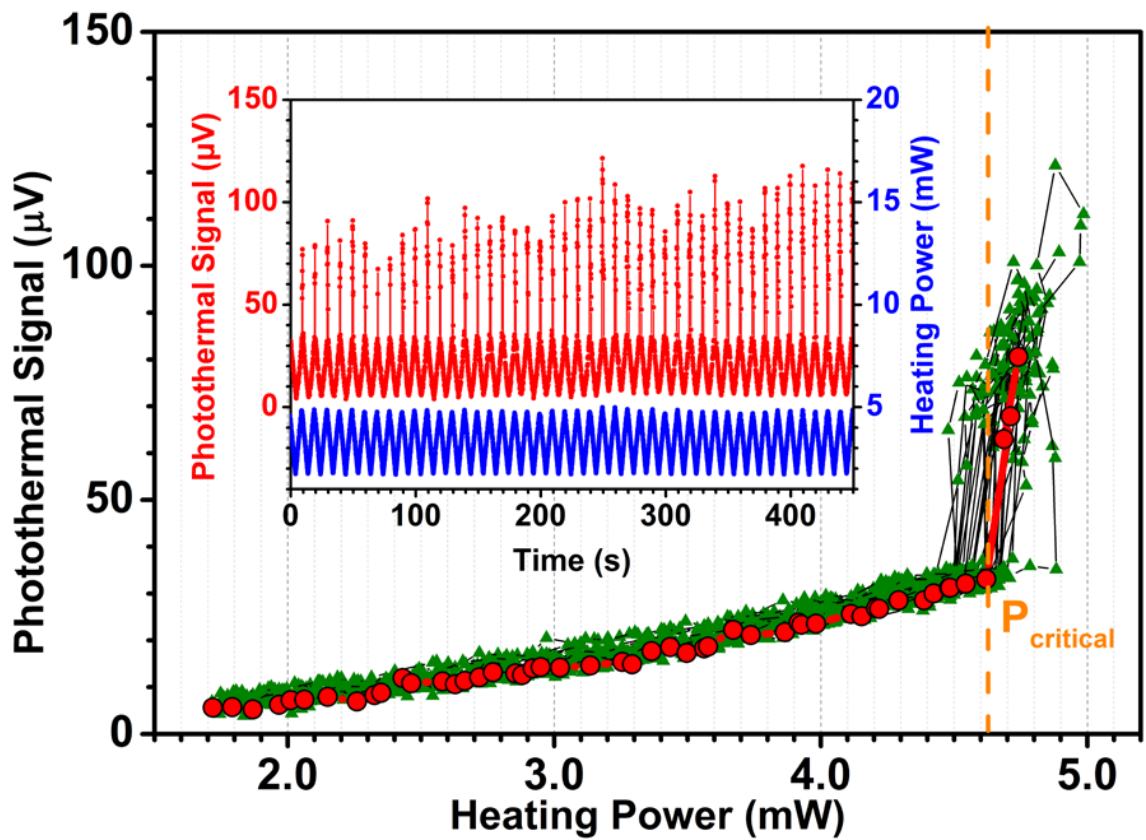

*Figure 3: Photothermal signal of a single gold nanosphere in water as the heating intensity is increased. Until a critical power of 4.6 mW (about 0.6 MW/cm$^2$) the signal increases smoothly as expected for heated liquid water. Above the critical power, the signal undergoes a sudden jump to a much higher value, with large fluctuations. Red dots: a typical example of a power sweep; green triangles: accumulated data from many*

*sweeps showing the dispersion in signal and in critical power. The data of each sweep are connected by a solid line. Insert: photothermal signal (red line, left scale) as a function of time while the heating power (blue line, right scale) is swept as a sawtooth function between values below and above the critical power.*

The model of Fig.1 shows that the particle temperature can rise by hundreds of K once the bubble forms. Such high temperatures may degrade the particle's shape and its contact area with the substrate. Therefore, we adjusted the intensity carefully to avoid damaging the particle. Moreover, we switched to a different liquid, n-pentane, which has a much lower boiling point than water, to further limit irreversible damage to the system and ensure reproducibility of the results. The physics of bubble formation and dynamics in pentane will thus serve as a model for bubble formation in water.

*3.2 Direct probe detection*

We thus directly detect the scattered probe intensity with the fast photodiode, following the bubble signal in real time. Figure 4 shows an example of a time trace recorded at 100 µW, just above the critical power (94 µW) in liquid pentane. The complex boiling trace of Fig. 4a appears as a succession of brief and violent events lasting some tens of ns only, separated by 500 ns on average. Such violent events are characteristic of explosive boiling, which is often observed in superheated liquids [27], and which can be suppressed with superhydrophobic coatings [28]. Herein, we use the word "explosion" to describe a rapid bubble expansion in the nanosecond time scale, similar to what is customary observed in pulsed heating experiments [1,6,10,29]. Note that these explosions occur at low Reynolds numbers, of the order of unity. The contraction or the decay part of the nanobubble signal resembles the collapse behavior of acoustically driven gas bubbles [1]. The signal-to-noise ratio is good enough to follow individual explosions (Fig. 4b), which present a rise time of about 14 ± 2 ns and a decay time of about 31 ± 7 ns, clearly longer than the detector's rise and fall times 5 and 16 ns. From these times and from the intensity and spatial dependence of the signal, we estimate the bubble radius to 100-200 nm (see Supplementary Information). We averaged hundreds of such events, synchronizing them with the rising edge of the explosion signal, and obtained the averaged profile of Fig.4c, which appears only slightly broadened by the averaging to a rise time of 18 ns. The decay part of the explosion signal presents a small but reproducible shoulder which will be discussed below (section 3.3). Beyond the main initial peak, the averaged trace shows further undulations at longer times, with average spacing of 500 ns, corresponding to the later explosions. They broaden because of the lack of exact periodicity.

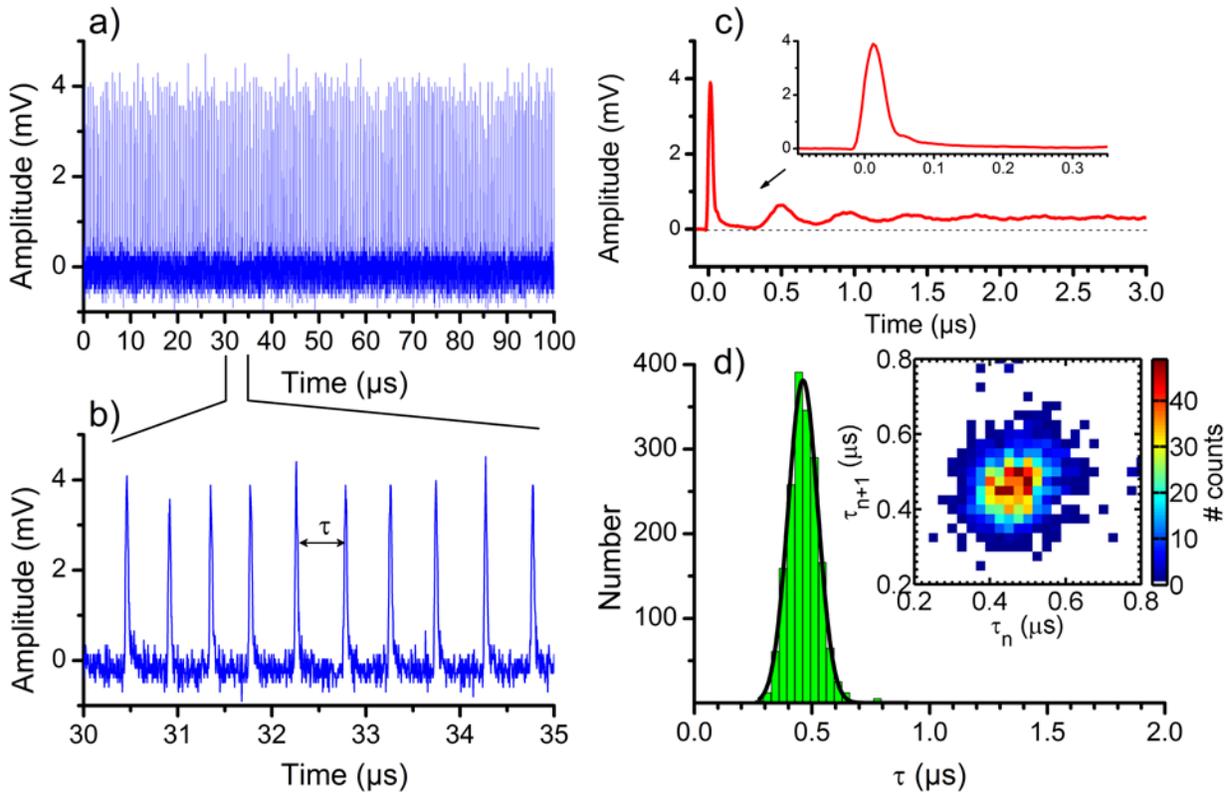

*Figure 4 Direct probe detection of nanobubbles under constant heating (100 µW) just above the threshold for boiling. Detector gain: ×1000, bandwidth: 10 MHz. (a) typical time trace of the scattered probe beam showing successive explosive nanobubble events. The probe power is fixed at 5 mW; (b) further zoom-in on one part of time trace (a); (c) Average of 3,123 explosions in time trace (a), taking the half maximum of the rising edge of the signal as the time reference. Insert: zoom-in on the main peak; (d) the histogram of delay times between two successive explosions. The solid line is a Gaussian fit. Insert: scatter plot of all pairs of consecutive delays, $\tau_n$ and $\tau_{n+1}$, from the trace in (a), showing no obvious correlation pattern.*

Taking a long trace with hundreds of bubble explosions, we can look at the distribution of inter-explosion delays, shown in Fig. 4d. No explosion is found to occur at less than 300 ns from the previous one. The distribution is well fitted by a Gaussian, with a maximum at about 500 ns for the conditions of Fig.4. We also present a scatter plot of the pairs of times between successive events ($\tau_n, \tau_{n+1}$) in a 2D diagram (Fig. 4d, insert). This plot appears compatible with a succession of inter-explosion times drawn at random from the Gaussian distribution of Fig. 4d. We thus conclude that the random noise causing jitter, or deviations from the average inter-explosion times, are uncorrelated between successive events. Similar jitter observations were reported in ref. [10], and attributed to randomness in the bubble nucleation process. In particular, the jitter in nanobubble dynamics is not caused by experimental imperfections such as laser intensity noise or focus drift. Note that heating intensity drifts can affect our measurements, as shown in the Supplementary Information.

We now propose and discuss a mechanism for this unexpected explosive boiling under continuous-wave heating conditions. As we saw in the introduction, to nucleate and grow, the nanobubble needs to overcome the Laplace pressure in addition to the ambient pressure. This only happens at 367 K for an 80-nm particle in pentane (483 K in water). Once the bubble starts to grow, however, the effective boiling temperature decreases because the Laplace pressure itself decreases. When a small part of the hot liquid vaporizes, it generates a first very thin vapor shell, which pushes the remaining hot liquid just outside the bubble. This hot liquid now becomes overheated with respect to the vapor in the bubble, because of the lowered Laplace pressure. It will then feed fresh steam into the bubble, further amplifying the expansion. We have estimated the energies involved in bubble growth (see Supplementary Information). We find three main contributions. Two of them are energy costs: i) the surface energy, which increases with bubble radius because both surface and surface tension increase, ii) the latent vaporization heat and internal vapor energy needed to expand the bubble. The third contribution is a source of energy, iii) the thermal energy stored in the overheated liquid layer. This heat can flow either to the cooler water layers at larger radii, or towards the bubble, helping its growth. For a final bubble radius of 140 nm, these contributions are – 1 fJ, -2.8 fJ, and 8.3 fJ, respectively, which indicates that bubble expansion liberates energy and is therefore thermodynamically favorable. Note that conduction through the liquid is fast enough to make this energy kinetically available during the expansion, as the diffusivity of heat in liquid pentane ($6\times10^{-8}$ $m^2$/s) corresponds to 8 nm in 1 ns. Once the excess thermal energy has been consumed into surface energy and latent heat, the bubble eventually reaches a maximum radius and shrinks back under the restoring forces of surface tension and vapor condensation. Indeed, at the maximum bubble radius, cooling of the thinned hot liquid layer by the outer cold liquid is very efficient. This explains that the bubble may disappear completely upon shrinking, as the returning cold bubble wall can condensate all the vapor. The cold liquid will have to be heated again by the nanoparticle during some hundreds of nanoseconds in our conditions before a new overheated layer is established and a new explosion can take place.

We note that a similar mechanism is at work in short-pulse experiments, where heat supplied by the hot nanoparticle first has to be conducted to the surrounding liquid before boiling can set in [13]. We can also compare our system to air bubbles in water. Those can reach very small minimal radii, with accordingly high temperatures and luminescence [30], followed by multiple after-bounces. In our case, however, the steam condensates until the bubble's surface hits the nanoparticle and the bubble disappears. Indeed, the data of Fig. 4b,c show that this contraction step is not followed by any clear after-bounces, apart from the small shoulder seen in Fig.4c at $t \approx 60$ ns, which will be discussed hereafter (see 3.3). The explosion repetition rate is mainly determined by the rate at which the overheated liquid layer, but it may also depend on the microscopic crossing of the nucleation barrier, as proposed recently by Nagashima et al. [10] in bubbles produced by Joule dissipation in a nanopore. Our analysis of time traces of Fig. 4 did not reveal any sign of a chaotic dynamics [31].

*3.3 Echo-triggered explosions*

Under finely tuned experimental conditions, a large fraction of explosion events are followed by after-pulses. Figure 5 shows an example observed with the particle studied in the experiments of Fig.4 but with a slightly lower heating power (97 µW), corresponding to an inter-explosion delay of about 1 µs. In contrast to Fig.4, the after-pulse of Fig.5 occurs at a well-defined delay of about 200 ns after the main explosion, which distinguishes it from the next explosion requiring a fully restored overheated layer. We propose that after-pulses are weaker explosions triggered by sound echoes of the main explosion, reflected from flat interfaces around the nanoparticle.

Two interfaces are possible candidates: i) the other side of the glass coverslip. With a thickness of 159 µm and a sound velocity of 5640 m/s [32], the echo arrives 56 ns after the explosion. This is precisely the delay of the shoulder seen in the decay of the bubble signal, both in Fig.4c and Fig.5c; ii) the interface between the immersion oil and the objective lens. The distance between the lens and the oil-glass interface is 100 µm (working distance according to the manufacturer). With a sound velocity of 1350 m/s in oil [33], this echo from the second interface should arrive about 204 ns after the main explosion, exactly as observed in Fig.5c. The after-pulse would thus be a second, weaker explosion triggered by the echo. It may feed on thermal energy left in the hot water layer after the first explosion or conducted from the hot particle. This remarkable phenomenon highlights the extreme sensitivity of a bubble to weak perturbations. Although the initial sound wave is attenuated by two transmissions through the glass-oil interface, reflection on an oil-glass interface and propagation as a spherical wave through 159 µm of glass and 100 µm of oil, this weak echo wave appears sufficient to trigger a measurable signal 56 ns after the first explosion and even a second explosion 204 ns later. In a few cases (see Supplementary Information), a second after-pulse follows the first one and can be attributed to an additional reflection.

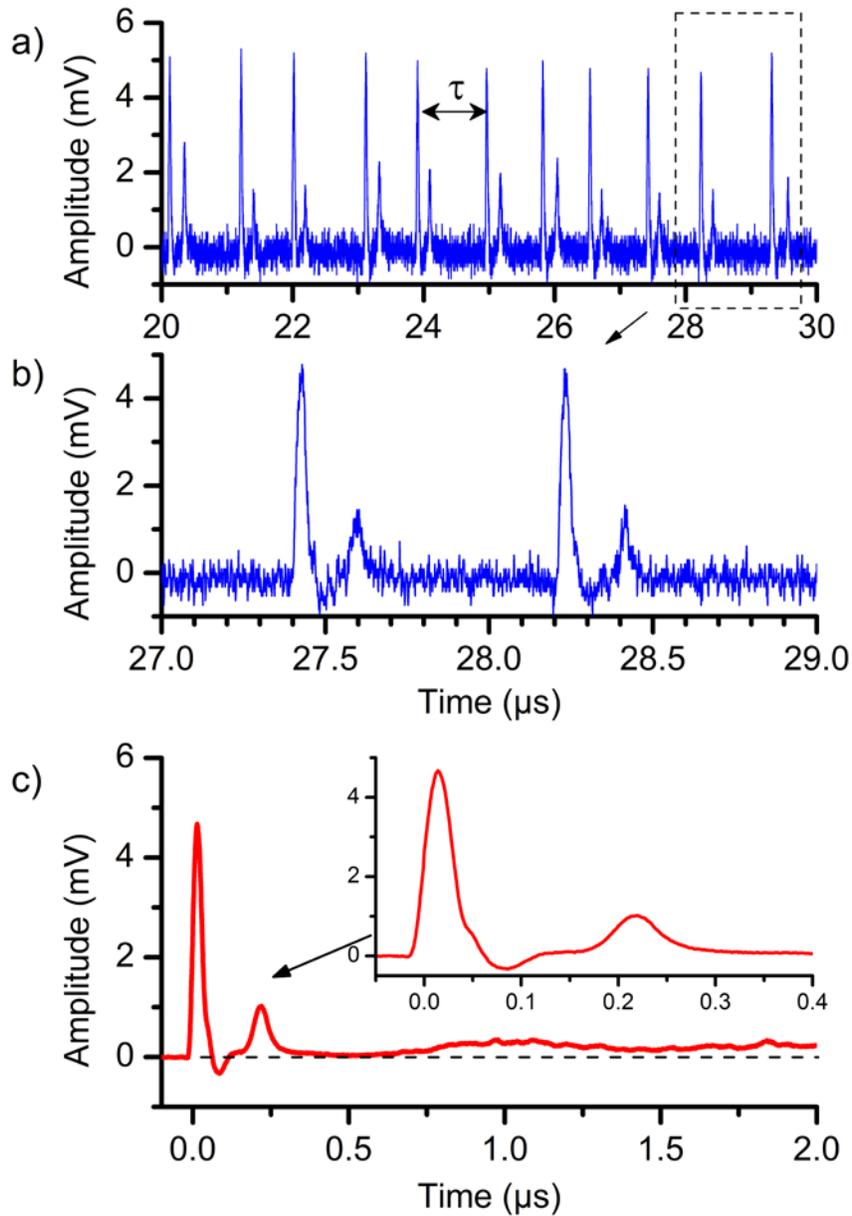

*Figure 5* Echo-triggered nanobubble explosions under finely-tuned constant heating power above the threshold. (a)The non-averaged raw time trace of scattered probe presenting echo-triggered bubbles. The probe power is fixed at 5 mW; (b) zoom-in time trace on the dash-block part in (a); (c) averaging signal of 1108 explosions in (a) in the same way as Fig. 4(c). Insert: zoom-in on the main peak of (c);

*3.4 Towards stabilization of a nanobubble*

The nanobubble around the particle could be stabilized by means of a suitable time profile of the laser intensity. The instable layer of overheated liquid around the nanoparticle makes it impossible to pass smoothly from regime I to regime II in Fig.1. However, the inverse process, in which the heating power is continuously decreased from a point in regime II does not generate any unstable situation. Indeed, the continuous presence of the liquid/vapor interface ensures that the two phases remain in equilibrium at all times, so that the bubble disappears in a continuous way when the heating power is reduced.

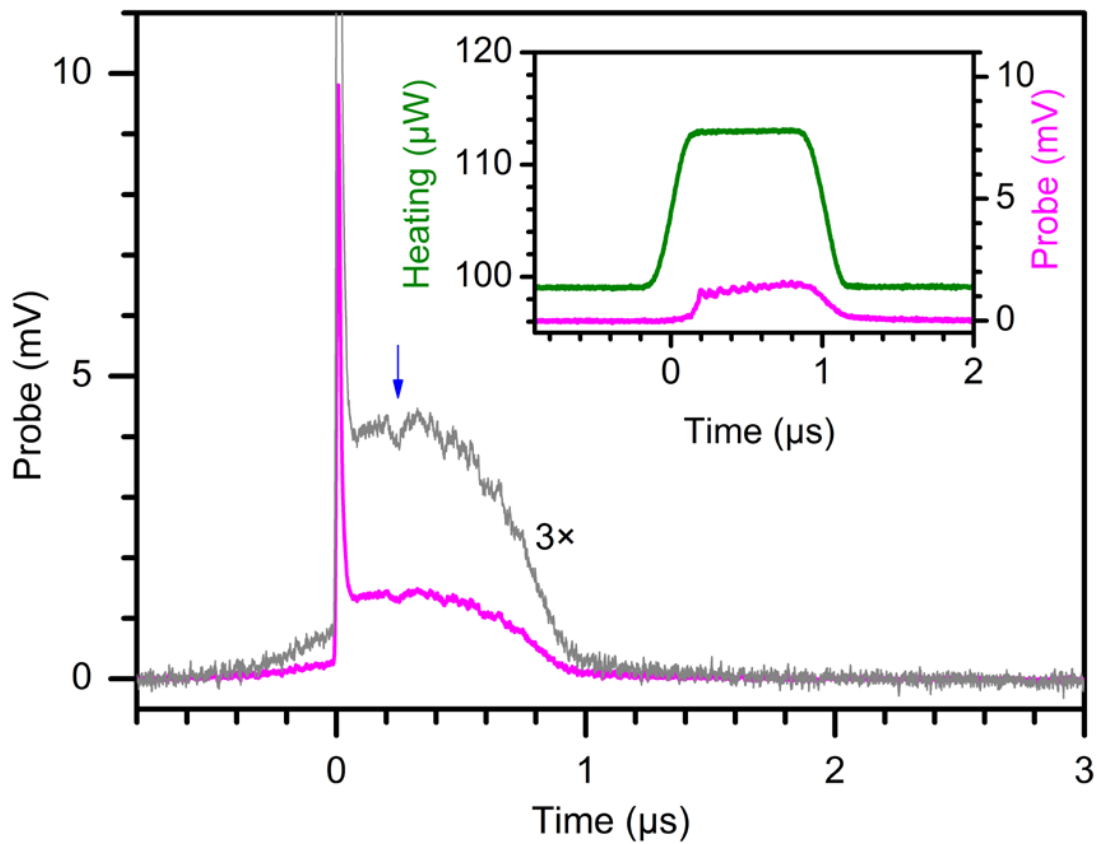

*Figure 6: Formation of a persistent nanobubble. The pink line shows the probe signal averaged over 200 explosive events following triggering by a raise in heating power. Detector gain: ✕ 10,000, bandwidth: 200 MHz. The events have been averaged by synchronizing at mid-rising edge of the probe transient signal. Insert: The averaged heating (green) and probe profile (pink), synchronized at mid-rising edge of the periodic heating pulse as delivered by the acousto-optic modulator. After explosive formation, the nanobubble persists for up to 800 ns. The heating beam is modulated by a block pulse profile with a frequency of 100 kHz and a duty cycle of 10% (1 µs on-time in a 10 µs period). Obviously, the explosive appearance of the nanobubble follows the rise in heating power with a jitter, washing out the explosion signature in the insert. A histogram of the jitter delays and single-shot time traces are displayed in the Supplementary Information.*

We therefore designed the following time profile of the laser power to lengthen the nanobubble's persistence time. We start just below the critical boiling power (0.96 $P_c$), then suddenly raise the heating intensity to a high value (1.1 $P_c$; in practice, due to the finite response time of our acousto-optic modulator, the rise in heating lasts about 100 ns). We then keep the intensity at this high level for a variable duration, 1 µs in the case of Fig.6, before reducing it back to the initial level. The result of this cycle for the scattered light is shown in Fig.6 together with the heating intensity profile.

Averaged probe signal traces following a raise in heating power are presented in Fig.6 (see a single-shot trace in the Supplementary Information). Again, the individual single-shot traces were average by synchronizing them on the mid-rising edge of the probe signal. The averaged trace clearly shows an initial explosion of about 30 ns duration and of lower amplitude than those in Figs. 4,5, followed by a plateau at a high scattering value. The probe signal after the explosion is much higher than before, when it was due to the liquid's temperature change alone. This high value indicates the presence of the bubble, and its persistence for as long as the heating power is kept at the high level. The bubble disappears as soon as the heating power is reduced. We therefore conclude that, in the few hundred ns following the explosion, the bubble reaches the steady-state extent discussed and calculated above. After the shrinking phase of the explosion, the bubble remains as a thin stable shell because enough heating power is provided, and the energy received by the vapor shell from the particle balances the energy lost by conduction to the cooler liquid outside. This experiment shows the feasibility of reaching and maintaining steady state II, once the barrier of bubble formation is passed. Much longer times than 1 µs could be achieved by optimizing the time profile and intensity stability of the heating power or by a proper, fast enough feedback mechanism from the scattered signal. A reproducible feature (shown with a blue arrow) appears on the trace of Fig.6, about 200 ns after the initial explosion. We assign this feature to the reaction of the nanobubble to the sound echo reflected by the oil-lens interface, and discussed in the previous section.

In some measurements with the same protocol in water, we observed self-oscillations of the nanobubble. Figure 7 shows a single-shot observation of a nanobubble formed in water upon a 1-µs raise of heating power. This bubble appears without an initial explosion and starts to oscillate after a few hundreds of ns. The oscillations damp out when the power is decreased again below the critical value. The oscillation period is roughly consistent to the time given by the Minnaert oscillation period of a bubble [34]. The possible mechanism of this self-oscillation is still unclear and requires additional investigation in the future.

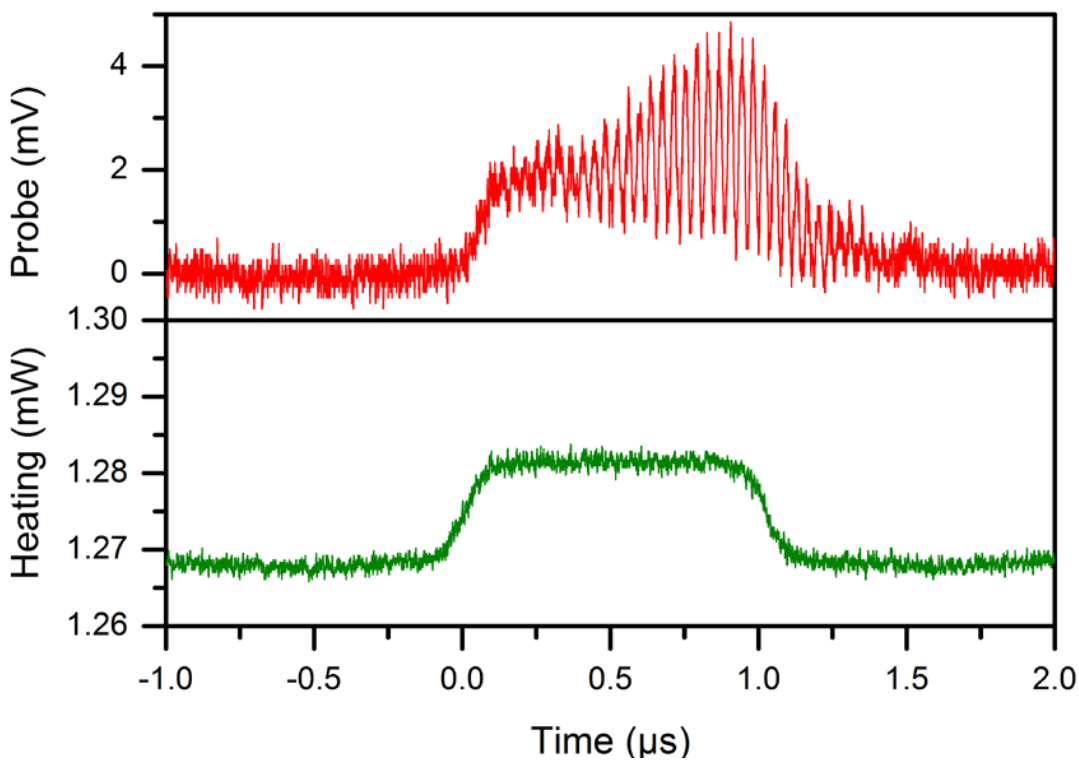

*Figure 7: A single-shot time trace showing the self-oscillating behavior of a nanobubble. Above: probe signal; below: heating profile. In this measurement, the gold particle is in pure water and on BK7 glass substrate. The duration of the heating "pulse" is 1 µs. The oscillation period of the bubble is about 30 ns.*

4 . Conclusion and outlook:

Boiling of a liquid around a heated metal nanoparticle can be controlled and detected with high sensitivity and fast time response. Even under continuous-wave heating, nanobubble formation is explosive. No after-bounce could be detected upon bubble shrinking, presumably because all kinetic energy is dissipated upon vapor condensation. The fast time response (less than 15 ns for the expansion and 20-30 ns for the contraction) could be used for all-optical light modulation with a bandwidth of about 100 MHz, several orders of magnitude faster than with liquid crystals [17]. Within a narrow range of heating power, the nanobubble becomes extremely sensitive to weak perturbations such as sound waves reflected from far-away interfaces. Acoustic wave fronts released in an initial explosion can trigger a new explosion, or lead to self-oscillations. We have shown that a steam nanobubble can be stabilized with a suitable heating intensity profile, and by controlling the heating laser power during the shrinking phase.

Our experimental results call for proper theoretical modeling. Compared to the inertial Rayleigh-Plesset theory and its refined versions including surface tension [35] and heat and mass transfer [3], the present system requires consideration of the thermodynamic and kinetic features of the liquid-gas phase transition [4] at nanometer scales. Moreover, the geometry of our experiment excludes spherical symmetry and calls for a full 3D model. Such a complex theory is well beyond the scope of the present work. Our results suggest using bubbles as nanoscale generators and detectors of acoustic waves, much as radars are used at macroscopic scales for electromagnetic waves.

Acknowledgements:

This work is supported by the Foundation for Fundamental Research on Matter (FOM) with funding from NWO. L.H. acknowledges the financial support of China Scholarship Council. Advice from Dr. P. Zijlstra and programming assistance of A. Carattino are kindly acknowledged.